\documentclass[twocolumn,aps,prd,showpacs,nofootinbib,superscriptaddress%
,preprintnumbers,floatfix,10pt]{revtex4-1} 

\usepackage{amsmath}
\usepackage{amsfonts}
\usepackage{amssymb}
\usepackage{epsfig}
\usepackage{graphicx}
\usepackage{color}
\usepackage{slashed}
\usepackage{epstopdf}
\usepackage{dsfont}
\usepackage{bigints}

\usepackage{float}

\usepackage{caption}
\usepackage{subcaption}

\usepackage{hyperref}

\newcommand{\I}{{i}}
\newcommand{\E}{\text{e}}
\newcommand{\tr}{\text{tr}}

\newcommand{\Tr}{\text{Tr}}
\newcommand{\Eqref}[1]{Eq.~\eqref{#1}}

\newcommand{\Nf}{N_{\mathrm{f}}}

\newcommand{\dir}{\slashed{\nabla}}
\newcommand{\kir}{k_{\text{IR}}}

\newcommand{\blamc}{\bar{\lambda}_{\text{cr}}}
\newcommand{\blamL}{\bar{\lambda}_\Lambda}

\newcommand{\psiBar}{\bar{\psi}}

\newcommand{\Det}{\text{Det}}

\begin{document}

\preprint{}

\title{A curvature bound from gravitational catalysis in thermal backgrounds} 

\author{Holger Gies}
\email{holger.gies@uni-jena.de}
\affiliation{Theoretisch-Physikalisches Institut, 
Abbe Center of Photonics, Friedrich Schiller University Jena, Max Wien 
Platz 1, 07743 Jena, Germany}
\affiliation{Helmholtz-Institut Jena, Fr\"obelstieg 3, D-07743 Jena, Germany}
\author{Abdol Sabor Salek}
\email{abdol.sabor.salek@uni-jena.de}
\affiliation{Theoretisch-Physikalisches Institut, 
Abbe Center of Photonics, Friedrich Schiller University Jena, Max Wien 
Platz 1, 07743 Jena, Germany}

\begin{abstract}
We investigate the phenomenon of gravitational catalysis, i.e.,
curvature-induced chiral symmetry breaking and fermion mass generation, at 
finite temperature. Using a scale-dependent analysis, we derive a thermal 
bound 
on the curvature of local patches of spacetime. This bound quantifies regions 
in parameter space that remain unaffected by gravitational catalysis and thus 
are compatible with the existence of light fermions as observed in Nature. 
While finite temperature generically relaxes the curvature bound, we observe 
a comparatively strong dependence of the phenomenon on the details of the 
curvature.  

Our bound can be applied to scenarios of quantum gravity, as any realistic 
candidate has to accommodate a sufficient number of light fermions. We argue 
that our bound therefore represents a test for quantum gravity scenarios: a 
suitably averaged spacetime in the (trans-)Planckian regime that satisfies our 
curvature bound does not induce correspondingly large Planckian fermion 
masses by gravitational catalysis. The temperature dependence derived in this 
work facilitates to follow the fate of gravitational catalysis during the thermal 
history of the (quantum) universe. 
In an application to the Asymptotic Safety scenario of quantum gravity, our 
bound translates into a temperature-dependent upper bound on the number of 
fermion flavors.  
  
\end{abstract}


\maketitle

\section{Introduction}
\label{intro}

Chiral symmetry breaking and fermion mass generation is a central feature of 
interacting fermions relevant for both the Higgs sector of the standard model as
well as QCD shaping 
many properties of matter in the universe. Whereas the long-range limit of 
gravity in the form of Einstein's general relativity is too weakly 
interacting to affect the status of chiral symmetry, gravity is expected to 
become more strongly interacting at or above the Planck scale. Whether or not 
gravity or its quantized form may exert a strong influence on the chiral 
features of fermions deserves to be studied. In fact, such an influence may 
even be used as an observational probe for scenarios of quantum gravity: as 
suggested in \cite{Eichhorn:2011pc}, viable scenarios of quantum gravity need 
to be compatible with the existence of light fermion as observed in Nature -- 
a requirement that has the potential to impose constraints or even rule out 
certain scenarios of quantum gravity.

It is reassuring to see that quantum fluctuations of the metric do not 
support the same kind of chiral-symmetry breaking mechanism as is triggered 
by spin-one gauge fields or Yukawa interactions with scalars 
\cite{Zanusso:2009bs,Eichhorn:2011pc,Meibohm:2016mkp,Eichhorn:2016esv,% 
Eichhorn:2016vvy,%
Eichhorn:2018nda,Catterall:2018dns,deBrito:2019epw,Daas:2020dyo}. For 
both latter cases, 
the gauge or Yukawa couplings simply have to increase beyond a certain 
threshold which renders chiral-symmetry breaking in these scenarios a rather 
universal strong-coupling feature. This is not so in metric quantum gravity. 

By contrast, gravity offers further mechanisms to trigger fermion mass 
generation which are generic to gravity in the sense that they proceed via 
the structure of spacetime itself. The most widely studied mechanism occurs 
on negatively curved spacetimes and can be summarized by \textit{gravitational 
catalysis} \cite{Ebert:2008pc}. It appears in a large variety of fermionic 
models 
\cite{Buchbinder:1989fz,Buchbinder:1989ah,Buchbinder:1989ah,%
Inagaki:1993ya, Sachs:1993ss,Elizalde:1993kb,Elizalde:1995kg,%
Kanemura:1995sx,Inagaki:1995bk,%
Inagaki:1996nb,Miele:1996rp,Vitale:1998wm,Inagaki:1997kz,Hashida:1999wb,% 
Gorbar:2007kd,Hayashi:2008bm,Inagaki:2010py,Gorbar:1999wa}, as it derives 
from a mechanism of dimensional reduction of the spectrum of the Dirac 
operator on hyperbolic spacetimes \cite{Gorbar:2008sp}. Another mechanism 
has recently been suggested and worked out in \cite{Hamada:2020mug}: in 
quantum gravity scenarios allowing for topology fluctuations, gravitational 
instantons can contribute to anomalous chiral symmetry breaking and thereby 
generate fermion masses potentially in conflict with observation. In 
combination with abelian gauge interactions, gravity can trigger also 
conventional symmetry breaking mechanisms, as demonstrated in 
\cite{deBrito:2020dta}.

In the present 
work, we further explore pure gravitational catalysis specifically by including 
the 
effects of finite temperature. Following an earlier zero-temperature analysis 
\cite{Gies:2018jnv,Martini:2019izh}, we study the phenomenon using a 
renormalization-group
(RG) inspired scale-dependent approach. The advantage is that we can monitor 
the RG relevance of chiral interactions in this way. In fact, gravitational catalysis 
can be connected with four-fermion operators becoming RG relevant driving the 
symmetry-breaking interactions to criticality \cite{Gies:2013dca}. This makes 
the analysis of gravitational catalysis in the context of quantum gravity 
scenarios more subtle: It is not sufficient to check, whether the long-range 
curvature of spacetime is compatible with the existence of light fermions (which 
obviously is the case). Moreover, the influence of spacetime curvature on the 
symmetry-breaking operators has to be checked during the whole course of the 
RG flow, specifically in the Planckian regime and beyond. Provided a notion of 
curvature exists in that regime, gravitational catalysis could be active and drive 
the symmetry-breaking operators beyond criticality. This would result in 
correspondingly heavy fermions removing light fermions from the observable 
long-range spectrum. 

This mechanism has been explored in \cite{Gies:2018jnv} which lead to the 
notion of curvature bounds: in order to guarantee that a given quantum gravity 
scenario is not affected by the problem of gravitational catalysis, 
the averaged curvature of a local patch of spacetime should not exceed a 
certain bound. So far, these bounds have been derived for Riemannian 
hyperbolic spacetimes such as $\mathbb{H}^D$ in general spacetime
dimensions $D$, cf. \cite{Gies:2018jnv}. 

In the present work, we generalize the analysis to $\mathbb{R} 
\otimes\mathbb{H}^{D-1}$ or $S^1\otimes\mathbb{H}^{D-1}$. The purpose is 
two-fold: first, this provides further information about the concrete 
dependence of the mechanism on the details of the averaged spacetime 
structure.

Second, this allows to monitor the influence of finite temperature on the 
mechanism. The latter is particularly relevant for studying the influence of 
gravitational catalysis in the course of the cosmological evolution. Indeed, 
our results provide evidence for a comparatively strong dependence of 
gravitational catalysis on the details of the background. At the same time, 
finite-temperature effects can significantly relax the curvature bounds -- in 
line with the expectation that thermal fluctuations drive the system towards 
the disordered symmetric phase.      

Our paper is organized as follows: Sect.~\ref{sec:framework} lays out
the general framework of our study in terms of a generic chiral fermion theory 
in curved spacetime, which we analyze in a local mean-field RG approach. 
The essential technical ingredient for taking the curved as well as 
thermal background into account, namely the heat kernel, is briefly described 
in Sect.~\ref{sec:heatkernel}.  
The curvature bounds for gravitational catalysis in a purely spatially 
hyperbolic spacetime with and without finite temperature are derived in Sect. 
~\ref{sec:1+3}. As an illustration, we apply these bounds from gravitational 
catalysis to the asymptotic-safety scenario for quantum gravity in Sect. 
~\ref{sec:AS}. In this application, our 
curvature bounds translate into an upper bound for the number of fermion 
degrees of freedom and zero temperature, and a combination of a bound on the 
fermion number and the temperature if considered within the context of a 
thermal history of the universe. We conclude in Sect.~\ref{sec:conc}.

\section{Chiral channel and effective potential}
\label{sec:framework}

In an RG picture, catalysis of chiral symmetry is triggered by 
four-fermion operators becoming RG relevant \cite{Gies:2013dca}. Considering 
$\Nf$ fermion flavors, we study the RG behavior of four-fermion operators with 
maximal chiral $U(\Nf)_\text{R} \times U(\Nf)_\text{L}$ symmetry as an example.

Operators with a lower degree of symmetry can be studied analogously. We focus 
on the so-called $(V)+(A)$ channel,
\begin{equation}
S_{\text{int}} \sim \int_x \left[\left( \psiBar^a \gamma_\mu \psi^a \right)^2 +
\left( 
\psiBar^a \gamma_\mu \I \gamma_5 \psi^a \right)^2 \right], \label{eq:Sint1} 
\end{equation}
which is one out of the two Fierz-independent local interaction terms of 
maximal symmetry \cite{Gies:2003dp}. It is Fierz equivalent to the 
scalar-pseudoscalar channel of the Nambu--Jona-Lasinio (NJL) model which, 
using the projectors 
\begin{align}
P_{\text{L}} = \frac{\mathds{1} - \gamma_5}{2}, \quad P_{\text{R}} = 
\frac{\mathds{1} + \gamma_5}{2},  \quad \mathds{1} = P_{\text{L}} + 
P_{\text{R}} 
\end{align}
onto left and right chiral components, can be re-arranged as
\begin{align}
\label{eq:SintpreHS}
S_{\text{int}}[\psiBar,\psi] = - 2 \int_x \bar{\lambda}  (\psiBar^a 
P_{\text{R}} \psi^b)(\psiBar^b P_{\text{L}} \psi^a).
\end{align}
Here we have introduced a (dimensionful) coupling constant $\bar\lambda$ 
parameterizing the strength of the chiral interaction. In the NJL model, this 
coupling is tuned beyond a critical value 
$\bar\lambda>\bar{\lambda}_{\text{cr}}$ triggering chiral symmetry breaking in 
terms of initial conditions. Incidentally, a thermal environment -- breaking 
spacetime symmetries explicitly -- allows for further sets of Fierz 
inequivalent interactions where spatial and temporal components of vector type 
channels are treated independently 
\cite{Braun:2017srn,Braun:2018bik,Braun:2019aow}. In the following, we ignore 
this potential splitting and concentrate on the NJL channel. Here, we always 
assume the initial condition to be subcritical such that this 
operator does not generate fermion 
masses on its own. 

Introducing a non-dynamical Hubbard-Stratonovich field $\phi$, the chiral 
channel can be rewritten in terms of a local Yukawa interaction,
\begin{align}
\label{eq:SintpostHS}
\mathcal{L}_{\text{int}}[\phi,\psiBar,\psi] = \psiBar^a \left[ P_{\text{L}} 
(\phi^\dagger)_{ab} + P_{\text{R}} \phi_{ab} \right] \psi^b + \frac{1}{2 
\bar{\lambda}} \tr(\phi^\dagger \phi). 
\end{align}
The equivalence between \Eqref{eq:SintpreHS} and 
\Eqref{eq:SintpostHS} becomes obvious with the aid of the equation of 
motion for the chiral matrix field,
\begin{align}
\phi_{ab} &= -2 \bar{\lambda} \psiBar^b P_{\text{L}} \psi^a \nonumber, \\
(\phi^\dagger)_{ab} &= -2 \bar{\lambda} \psiBar^b P_{\text{R}} \psi^a.
\end{align}
This scalar field, in fact, serves as an order parameter for the status of 
chiral symmetry. E.g., assuming a diagonalizable expectation value in flavor 
space, $ \phi_{ab}  = \phi_0 \delta_{ab}$ with $\phi_0>0$ being homogeneous in 
spacetime, the chiral group breaks to a residual vector symmetry 
similar to QCD-like theories, and all fermions acquire masses of order 
$\phi_0$. Including a fermion kinetic term, the action reads
\begin{align}
S[\phi_0,\psiBar,\psi] = \bigintsss_x & \Big\{ \psiBar \left( \slashed{\nabla} 
+ \phi_0 \right) \psi +    \frac{1}{2 \bar{\lambda}} \Nf (\phi_0)^2
 \Big\}.
\end{align}
Integrating out the fermion degrees of freedom, we obtain a mean-field 
expression for the effective potential of the order parameter
\begin{align}
\tilde{U}(\phi_0) &= \frac{\Nf}{2 \bar{\lambda}}  (\phi_0)^2 - \Nf \log \Det_x 
(\slashed{\nabla} + \phi_0) \nonumber \\
% &= \frac{\Nf}{2 \bar{\lambda}}  (\phi_0)^2 - \dfrac{\Nf}{2} \log \Det_x 
% (-\slashed{\nabla}^2 + \phi_0^2) \nonumber \\
&= \frac{\Nf}{2 \bar{\lambda}}  (\phi_0)^2 - \dfrac{\Nf}{2} \Tr_x \log 
(-\slashed{\nabla}^2+ \phi_0^2),
\end{align}
where we have used the $\gamma_5$-hermiticity of the covariant Dirac operator 
in 
the last step. It is convenient to introduce the Fock-Schwinger propertime
representation, 
\begin{align}
\label{effUtildeSPT}
\tilde{U}(\phi_0) = \frac{\Nf}{2 \bar{\lambda}}  (\phi_0)^2 + \dfrac{\Nf}{2} 
\bigintsss_0^\infty \dfrac{ds}{s}\E^{-\phi_0^2s} \Tr_x \
\E^{\slashed{\nabla}^2s},
\end{align}
in order to arrive at the heat-kernel trace for the present differential operator of 
interest:
\begin{align}
\Tr_x \ \E^{\slashed{\nabla}^2s} = \Tr_x \ K(x,x';s) =: K_D(s).
\end{align}
The heat kernel $K(x,x';s)$ satisfies a modified heat flow equation with the 
following boundary conditions
\begin{align}
\dfrac{\partial}{\partial s} K = \slashed{\nabla}^2 K, \quad \lim_{s
\rightarrow 
0^+} K(x,x';s) = \dfrac{\delta(x-x')}{\sqrt{g}}. 
\end{align}
The propertime representation is not only useful to evaluate the functional 
trace of the heat kernel on curved spacetimes, but also allows to regularize 
this fermionic fluctuation contribution in a scale-dependent and 
spin-base-invariant \cite{Gies:2013noa} fashion: 
contributions from the infrared (IR) modes of the fermionic spectrum contribute
predominantly to the large-$s$ part of the propertime integral. Hence, these 
modes can be IR regularized 
by insertion of a regulator function $f_k$, 
\begin{align}
\label{eq:fk}
f_k = \E^{-(k^2 s)^p},
\end{align}
into the propertime integral. The parameter $p>0$ specifies the renormalization
scheme and $k$ corresponds to an IR regularization scale for the eigenvalues 
of the squared Dirac operator. For $p \rightarrow \infty$, all long range 
contributions are sharply cut off at the scale $s>1/k^2$. For finite values 
of $p$, the regularization scale is smeared out. In the limit $k \rightarrow 0$, the 
RG insertion factor becomes the identity, and the regularization is thus removed. 
Starting at an ultraviolet (UV) scale $k=\Lambda$ with the bare potential $
\tilde{U}_\Lambda$ , the potential in the IR at $\kir$ can be computed by 
\begin{align}
\label{eq:flatbc}
\tilde{U}_{\kir} =\tilde{U}_\Lambda - \bigintsss_{\kir}^\Lambda dk \ \partial_k
\tilde{U}_k, \quad \tilde{U}_\Lambda = \frac{\Nf}{2 \bar{\lambda}_\Lambda}  
\phi_0^2 , \quad \bar{\lambda}_\Lambda := \bar{\lambda}. 
\end{align}
At intermediate scales $k$, the scale-dependent effective potential $\tilde{U}_k$ 
satisfies the flow equation
\begin{align}
\label{eq:FinalEffPotEq}
\partial_k \tilde{U}_k = \dfrac{\Nf}{2}  \bigintsss_0^\infty 
\dfrac{ds}{s}\E^{-\phi_0^2s} (\partial_k f_k ) K_D(s). 
\end{align}
The advantage of performing the integral over the Schwinger propertime $s$ 
first is that the cutoff $\Lambda$ controls the UV divergences and thus 
assists to identify and fix counter terms for the corresponding relevant and 
marginal operators.

\section{Heat kernels}
\label{sec:heatkernel}

Aiming at an analysis of the scale-dependent effective potential of 
Eq.~\eqref{eq:FinalEffPotEq}, the information about the spacetime structure 
enters via the heat-kernel trace $K_D(s)$. As we are interested in the 
mechanism of 
gravitational catalysis and the influence of finite temperature, we focus on 
spacetimes that feature a sufficient amount of negative curvature and allow for
a simple use of thermal field theory in imaginary-time formalism. Therefore, a 
natural choice is $S^1 \otimes H^d$ with a compactified (Euclidean) time and 
the spatial part corresponding to a maximally symmetric hyperboloid with 
negative spatial curvature. The decompactified limit then corresponds to the 
zero-temperature case $\mathbb{R} \otimes H^d$ with a flat time direction. 

It is important to emphasize that we do not at all consider these spacetimes as
physical descriptions of the large-scale structure of the universe. By means of
our scale-dependent analysis, we focus on effective properties of quantum 
spacetime, say, in the trans-Planckian regime. Here, nothing specific is known 
about the microscopic spacetime structure. Hence, our choice of spacetime can 
be considered as a proxy for a possible structure of local patches of spacetime in 
that short-distance regime of quantum gravity. For the product manifolds 
considered here, the square of the Dirac operator can be decomposed as
\begin{align}
\dir^2_D = (\partial_0)^2 + \dir^2_d, \qquad D=d+1.
\end{align}
Correspondingly, the heat-kernel trace factorizes,
\begin{equation}
K_D(s) = \Tr_t \ \E^{(\partial_0)^2s}  \cdot \Tr_x \ \E^{\dir^2_d s}
= K_t(s) \cdot K_d(s),
\label{eq:hktfactor}
\end{equation}
Let us first discuss the spatial part $K_d(s)$ for which an analytical result 
exists and has been worked out for general dimensions $d$ 
\cite{Camporesi:1992tm}. Focusing in this work on $d=3$-dimensional space, 
the result is particularly simple:
\begin{align}
K_{d=3}(s) = \dfrac{1}{(2 \sqrt{\pi s})^3}  \left( 1+\dfrac{1}{2}\kappa^2 s 
\right),
\label{eq:Kd3}
\end{align}
which holds for an arbitrary curvature parameter
\begin{align}
\kappa^2 = -\dfrac{R}{d(d-1)} = -\dfrac{R}{6} > 0.
\end{align}
The temporal part depends on the circumference $\beta=\frac{1}{T}$ of the 
Euclidean time $S^1$. Using anti-periodic boundary conditions for the fermionic
fields, the trace if performed in momentum space runs over Matsubara 
frequencies $\omega_n= 2\pi T(n+1/2)$, yielding
\begin{equation}
K_t(s) = T \sum_{n=-\infty}^{\infty} \E^{-\omega_n^2s} = T\vartheta_2\left( 0, 
\E^{-(2\pi T)^2 s} \right).
\label{eq:matsubara}
\end{equation}
Here we encounter the Jacobi theta function $\vartheta_2(z,q)$.  For our 
purposes, a Poisson resummation connecting $\vartheta_2$ to  $\vartheta_3$ is 
useful for later numerical evaluation. It also gives direct access to analyic studies 
of the low temperature-limit implying the decompactification 
$S^1 \to \mathbb{R}$ of the Euclidean time direction,
\begin{align}
\label{eq:lowTExp}
K_t(s) &= T \vartheta_2\left( 0, \E^{-(2\pi T)^2 s} \right) \nonumber \\
&= \dfrac{\sqrt{\pi}}{\sqrt{(2\pi)^2 s}} \vartheta_3\left( \dfrac{\pi}{2}, 
\E^{-\frac{\pi^2}{(2\pi T)^2 s}} \right) \nonumber \\
&= \dfrac{1}{ \sqrt{4 \pi s}} \left[ 1 - 2\ \E^{-\frac{1}{4T^2s}} +\mathcal{O} 
\left( \left( \E^{-(4T^2s)^{-1}} \right)^2 \right) \right] \nonumber .
\end{align}
Here, we obtain the standard zero-temperature results $K_t(s) = 1/\sqrt{4\pi 
s}$ for a fully decompactified temporal direction.

% 
% \coled{I think the following part is no longer needed, may be deleted at some 
% point}
% 
% For later use, we also quote the high-temperature limit of the temporal 
% heat-kernel trace, corresponding to the lowest order $n=0,-1$ terms of the 
% Matsubara sum in Eq.~\eqref{eq:matsubara},
% %
% \begin{equation}
% \label{eq:highTExp}
% K_t(s) = 2  T \E^{-\pi^2T^2s} + \dots
% \end{equation}
% %
% with the higher-order terms being even more exponentially suppressed than the 
% leading one for $T^2 s\gg 1$. This exponential suppression signifies the 
% dimensional reduction of the high-temperature limit to an 
% effectively $d=3$ dimensional space.

\section{Curvature Bounds}
\label{sec:1+3}

We are now in a position to derive bounds on the curvature parameter that 
characterize the parameter space free of gravitational catalysis. For this, we 
follow the reasoning of \cite{Gies:2018jnv} and monitor the possible 
occurrence of nontrivial minima of the effective potential for the chiral order
parameter $\phi_0$. In addition to the divergencies associated with matter 
operators to be renormalized, see next subsection, the effective potential 
$\tilde{U}(\phi_0)$ displayed, e.g., in Eq.~\eqref{effUtildeSPT}, also contains
a divergent zero-point energy, which we subtract by defining
\begin{equation}
 U(\phi_0)= \tilde{U}(\phi_0) - \tilde{U}(0),
\label{eq:zeropointsub}
\end{equation}
such that $U(0)=0$ is fixed at the origin in field space \cite{Vassilevich:2003xt}. A possible mixing of 
the subtraction terms with the cosmological-constant term is not considered in 
this work; we assume the -- possibly scale-dependent -- behavior of the 
cosmological constant to be provided by a given quantum gravity scenario 
(including matter backreactions).

\subsection{Curvature Bounds at Zero Temperature}

Let us first work out the renormalization of the effective 
potential, identifying all free parameters by accordingly fixing the required 
renormalization counter terms. Using the preceding results for the heat-kernel 
traces, the zero-temperature effective potential of  Eq.~\eqref{effUtildeSPT} 
upon IR regularization \eqref{eq:fk} and zero-point subtraction 
\eqref{eq:zeropointsub} reads
\begin{align}
\label{eq:Pot0T}
U_k =& \frac{\Nf}{2 \bar{\lambda}}  \phi_0^2 \nonumber\\
& + \dfrac{\Nf}{2 (4 
\pi)^2}  \bigintsss_0^\infty \dfrac{ds}{s^3} f_k \ \left( \E^{-\phi_0^2s} - 1 
\right)   \ \left( 1+\dfrac{1}{2}\kappa^2 s\right).
\end{align}
A power-counting analysis reveals the occurrence of a quadratic divergence for 
the $\phi_0^2$ operator and two logarithmic divergences for the $\phi_0^4$ and 
$\phi_0^2 R$ operators, respectively.

As a sufficient criterion for the occurrence of chiral symmetry breaking, we 
specifically monitor the sign of the $\phi_0^2$ term in the Taylor 
expansion of the effective potential. If this sign turns negative, the 
$\phi_0^4$ operator cannot inhibit chiral symmetry breaking. For the curvature 
bound derived below, the $\phi_0^4$ operator is thus not relevant; from here 
on, we assume it to be properly renormalized such that the coupling has some 
finite value at the scale $k$ at which we consider the theory. We add that the 
sign criterion of the $\phi_0^2$ is not a necessary criterion for chiral 
symmetry breaking, as first-order-type transitions to a broken phase could go 
along with a positive $\phi_0^2$ term. We ignore this option in the 
following; if it was realized our curvature bound would even get stronger. 

The remaining divergences can conveniently be identified by using the flow 
equation \eqref{eq:FinalEffPotEq}, inserting the regulator \eqref{eq:fk} and 
expanding in $\phi_0$. To leading order, we obtain
\begin{align}
\partial_k U_k = \dfrac{k \Nf \phi_0^2}{2 (4 \pi)^2} \left[ 2 \ \Gamma 
\left(1-\dfrac{1}{p} \right) + \frac{\kappa^2}{k^2} \right] + 
\mathcal{O}(\phi_0^4) .
\end{align}
Here, we observe a divergence for the case of a regularization parameter $p=1$.

This is expected, as this value would correspond to a mass-type Callan-Symanzik
regularization scheme which is known to be insufficient for an adequate 
suppression of UV modes in 4 dimensions. In order not to be affected by 
this artificial divergence from the regulator, we suggest 
to use schemes with $p\geq 2$. 

Next, we integrate the flow from an IR scale $\kir$ to a UV scale $\Lambda$, 
using for the UV boundary condition not only the flat space expression as in 
Eq.~\eqref{eq:flatbc}, but also including a possible scalar-curvature counter 
term,
\begin{equation}
 U_\Lambda(\phi_0)= \frac{\Nf}{2 \bar{\lambda}_\Lambda} \phi_0^2 + \Nf 
\xi_\Lambda \phi_0^2 R,
\label{eq:counterU}
\end{equation}
with a UV coupling $\xi_\Lambda$. The resulting effective potential at $k=\kir$
then reads up to order $\phi_0^2$ and ignoring terms of order 
$\mathcal{O}(1/\Lambda)$:
\begin{align}
\label{eq:PotUzeroT}
U_{\kir} =& -\dfrac{\Nf \phi_0^2}{2} \left( \dfrac{1}{\blamc} - 
\dfrac{1}{\blamL}- \dfrac{\kir^2}{16 \pi^2} \Gamma \left(1-\dfrac{1}{p} \right)
\right) \nonumber \\
&-6 \Nf \xi_{\kir} \phi_0^2 \kappa^2+ \mathcal{O}(\phi_0^4) .
\end{align}
Here, we have introduced the (scheme-dependent) critical coupling of the chiral
channel
\begin{align}
\label{eq:lamcrT0}
\blamc = \dfrac{16\pi^2}{\Lambda^2 \Gamma \left(1-\dfrac{1}{p} \right) },
\end{align}
and defined the finite scalar-curvature coupling at the scale $\kir$ as 
\begin{align}
\xi_{\kir}  = \xi_\Lambda + \dfrac{1}{12(4\pi)^2} \log\left( 
\dfrac{\Lambda}{\kir} \right).
\end{align}
Equation \eqref{eq:PotUzeroT} can be interpreted as follows: the first line 
contains the information about the symmetry status in flat spacetime.
In a subcritical regime, e.g. $\blamL < \blamc$, the mass-like term
remains positive for zero curvature, indicating that the origin, $
\phi_0= 0$, is a local minimum of the potential (in fact, it is also a
global one); hence the system is in the disordered phase and the fermion mass 
remains zero. In the supercritical regime however, e.g. $\blamL > \blamc$, the 
mass-like term in the first line can become negative for decreasing $\kir$
resulting in a nontrivial minimum $\phi_0^2 > 0$ in the long-range limit. This 
implies chiral symmetry breaking and fermion mass generation in flat spacetime. 
Now, the second line of Eq.~\eqref{eq:PotUzeroT} contains the curvature 
contributions resulting from the hyperbolically curved space. Assuming 
$\xi_{\kir}$ to be positive, the prefactor of this second term is negative and can 
therefore cause chiral symmetry breaking depending on the magnitude of the 
terms in the first line. 
Of course, we assume the fermionic self-interactions to be subcritical, 
otherwise the system would be in an NJL-like phase which does not
conform with the low-mass scale of the standard-model fermions.
While finite values of $\blamL$ are expected to be generated by gauge
and Yukawa interactions, we use the following simple estimate for the first line of 
Eq.~\eqref{eq:PotUzeroT}:
\begin{align}
-\dfrac{\phi_0^2}{2} \left( \dfrac{1}{\blamc} - \dfrac{1}{\blamL}- 
\dfrac{\kir^2}{16 \pi^2} \Gamma \left(1-\frac{1}{p} \right) \right) & 
\nonumber \\
\geq & \phi_0^2\dfrac{\kir^2}{32 \pi^2} \Gamma \left(1-\dfrac{1}{p} 
\right).
\label{eq:EstBoundT0}
\end{align}
Comparing this to the curvature-dependent contribution $\sim \xi_{\kir}$, we 
conclude that gravitational catalysis does not occur, if the ratio of the 
curvature of local patches of spacetime to the energy scale satisfies
\begin{align}
\label{eq:BoundT0}
\dfrac{\kappa^2}{\kir^2} \leq \dfrac{\Gamma \left(1-\dfrac{1}{p} 
\right)}{192\pi^2 \xi_{\kir}}.
\end{align}
Any finite value of the fermionic self-interaction $\blamL$ at the high 
scale would even strengthen the bound. 
We observe an apparent explicit scheme dependence of 
our bound through the regularization parameter $p$. For 
the region $2\leq p<\infty$, this dependence is rather mild, since $1 < 
\Gamma(1- \frac{1}{p}) \leq \sqrt{\pi}$. However, it should be noted that also the 
left-hand side carries an implicit scheme dependence, since the
dimensionless ratio of curvature -- which we consider as an effective
curvature of spacetime patches -- and the IR scale $\kir$ depends on 
the details of the spacetime averaging procedure. As the latter, if done 
explicitly, would go hand in hand with the average over the fermionic
fluctuations on various length scales, we expect the existence of 
such a bound as in Eq.~\eqref{eq:BoundT0} to have a universal meaning. 
We take the residual $p$ dependence of \eqref{eq:BoundT0} as a
measure for our ignorance of the details of the averaging process.
It is instructive to compare this result for the $\mathbb{R}\otimes 
H^3$ background with the corresponding bound for the maximally 
symmetric case $H^4$.
Here, the heat-kernel trace is a nonpolynomial function of the
curvature leading to an integral representation of the curvature bound 
\cite{Gies:2018jnv}. For the purpose of the present discussion, we use 
the simple analytic approximation also given in \cite{Gies:2018jnv}:
\begin{equation}
H^4:\quad \frac{\kappa^3}{\kir^3}+\frac{4}{3}\frac{\pi^{\frac{5}{2}}}{
\Gamma\Big(1+\frac { 1 }{2p}\Big)}\xi_{\kir}\frac{\kappa^2}{\kir^2}\leq
\frac{\sqrt{\pi}}{2}\frac{\Gamma\Big(1-\frac{1}{p}\Big)}{\Gamma\Big(1+\frac{1}{
2p}\Big)}\,.
\label{eq:boundApprox4dim}
\end{equation}
Apart from numerical factors, the main difference arises from the
first term $\sim \kappa^3$ in the $H^4$ case which is 
present independently of the marginal scalar-curvature coupling $\sim 
\xi$. Though the curvature bound itself does depend on the precise 
value of $\xi_{\kir}$ also in $H^4$, there is a meaningful bound for 
any value of, say, $\xi_{\kir} \sim \mathcal{O}(1)$ with $\xi_{\kir}=0$
being a legitimate choice. 
This is not the case for our present result \eqref{eq:BoundT0} for the 
bound which depends strongly on $\xi_{\kir}$, yielding no meaningful 
result for $\xi_{\kir}=0$.  The
reason for this strong dependence lies in the fact that the heat-kernel 
trace on $H^3$ has the particularly simple polynomial form given in 
Eq.~\eqref{eq:Kd3}, the contribution of which to the effective potential 
can be fully absorbed in the renormalization of the marginal 
scalar-curvature coupling $\xi$. 

We draw the following conclusions from this observation: first, this 
strong qualitative and quantitative difference between the curvature 
bounds of two example spacetimes with negative curvature
demonstrates that the details of the average spacetime structure in the 
(trans-)Planckian regime of quantum gravity can take a strong
influence on the presence or absence of gravitational catalysis. If a 
bound derived for one case is satisfied it may still be violated in 
another case. Since we have little access to general
knowledge about the average spacetime structure in this short-distance regime 
where spacetime itself is expected to be strongly
fluctuating, the exclusion of gravitational catalysis in order to reach
compatibility with the existence of light fermions can thus be decisive
criterion for the viability of a quantum gravity scenario. 

Second, in addition to information about the averaged spacetime structure 
of local spacetime patches, a quantum gravity (plus matter) scenario has to 
provide also a prediction of the scalar curvature coupling $\xi$ in order to 
test for gravitational catalysis. Since the scalar field in the present 
analysis arises from fermion interactions which may arise predominantly from 
classically scale-invariant gauge interactions, the use of a conformally 
coupled scalar field is a reasonable first guess.

\subsection{Curvature Bounds at Finite Temperature}

\begin{figure}[H]
\center
\includegraphics[width= \columnwidth]{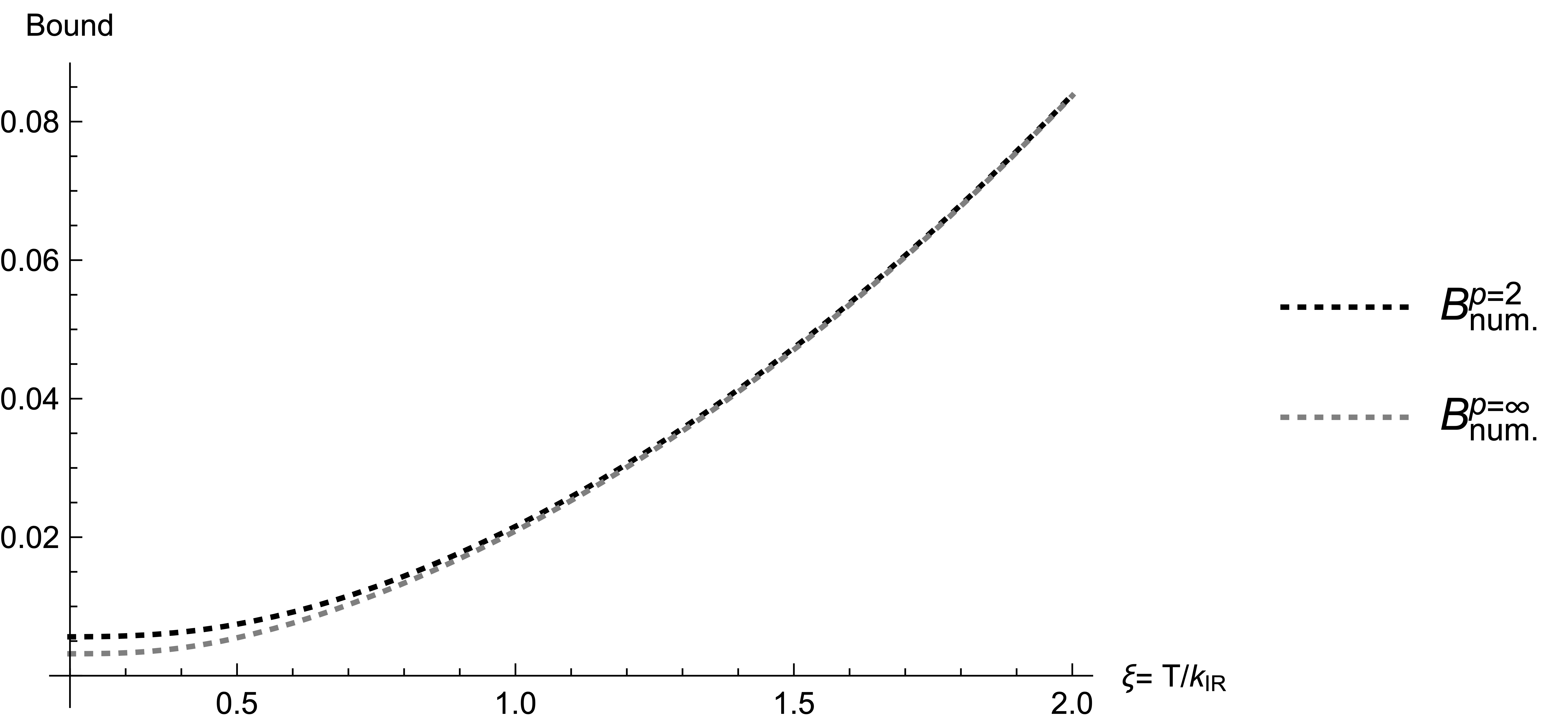}
\caption{\label{fig:nump1Inf} Numerical result for 
the curvature bound $B^p(\zeta)$ of Eq.~\eqref{eq:Boundpz} as a function of
the rescaled temperature $\zeta=T/\kir$ for regularization-scheme parameters 
$p=2$ and $p=\infty$, respectively. The comparatively mild scheme 
dependence at zero temperature even weakens for increasing temperature.}
\end{figure}

\begin{figure}[H]
\begin{center}
\includegraphics[width= \columnwidth]{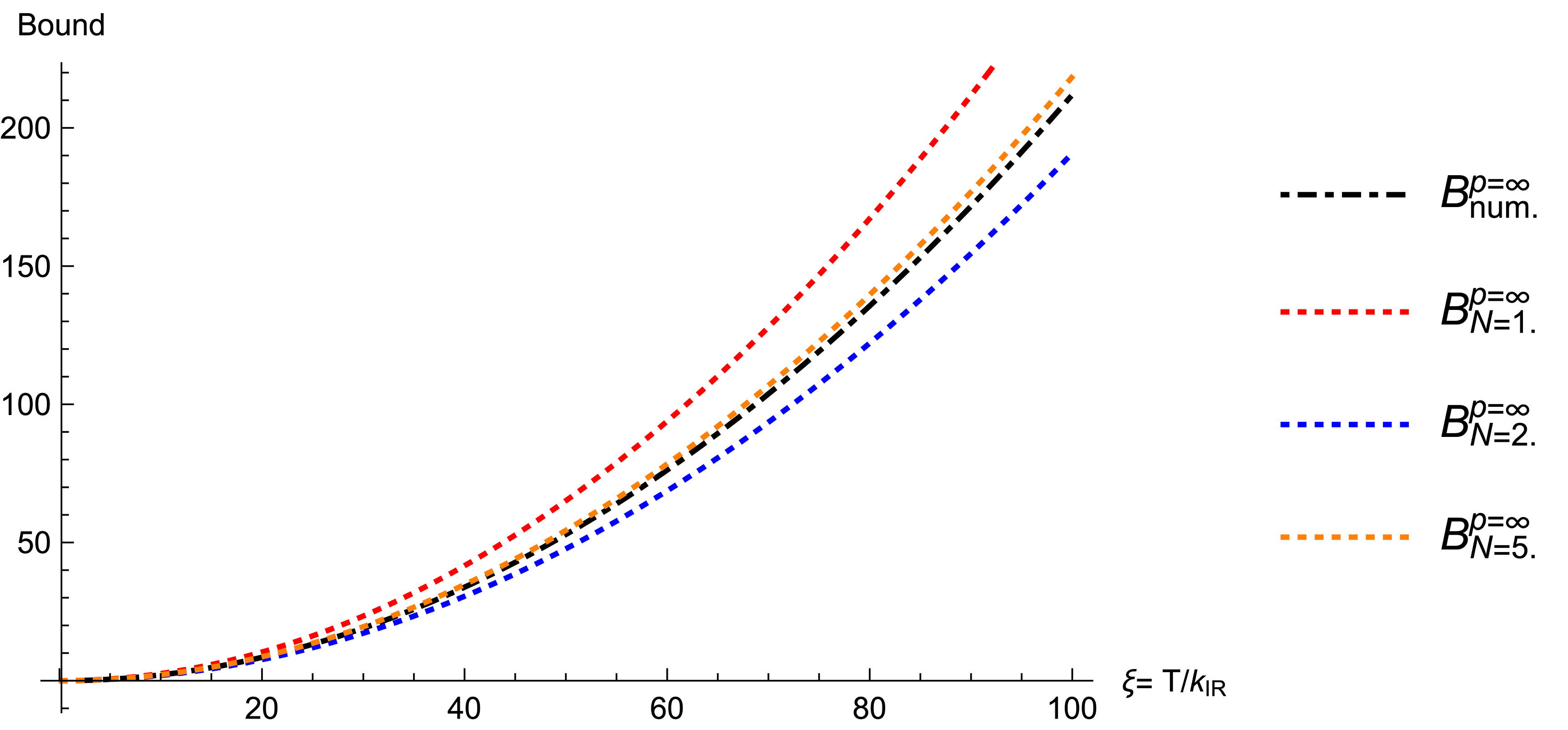}
\end{center}
\caption{\label{fig:numpinftyO135} Numerical result for the curvature bound 
$B^{p\to\infty}(\zeta)$ of Eq.~\eqref{eq:Boundpz} as a function of the rescaled 
temperature $\zeta=T/\kir$ in comparison with the analytical estimates of 
Eqs.~\eqref{eq:Apzeta} \eqref{eq:Cpzeta} for increasing truncations $N$. Even 
for large values of $\zeta$, the analytical estimates approach the full result
rather rapidly. For large temperatures, the curvature bound increases $\sim
\zeta^2$.}
\end{figure}

As in the zero-temperature case, we now derive curvature bounds from the 
effective potential for the chiral order parameter. For this, we write the 
regularized effective potential as
\begin{equation}
 U_k^T=U_k+\Delta_T U_k,
 \label{eq:DeltaT}
\end{equation}
where $U_k$ denotes the zero-temperature part, cf. Eq.~\eqref{eq:Pot0T}, and 
$\Delta_T U_k$ is the thermal correction satisfying $\Delta_{T=0} U_k=0$.
Based on the heat-kernel traces, this thermal part can be written as
\begin{eqnarray}
\Delta_T U_k&=& \dfrac{\Nf}{2 (4 
\pi)^2}  \bigintsss_0^\infty \dfrac{ds}{s^3} f_k \ \left( \E^{-\phi_0^2s} - 1 
\right)   \ \left( 1+\dfrac{1}{2}\kappa^2 s\right) \nonumber\\
&& \qquad \times \left[\vartheta_3\left( \dfrac{\pi}{2}, 
\E^{-\frac{\pi^2}{(2\pi T)^2 s}} \right) - 1\right].
\label{eq:DeltaTU}
\end{eqnarray}
Since the presence of finite temperature does not modify the UV behavior of the
theory, this expression is already finite. No further counterterms are required,
and we consider all physical parameters to be fixed by the $T=0$ 
renormalization conditions. After the substitution $\tilde{s}=\kir^2 s$, the
thermal correction to the effective potential up to quadratic order in $\phi_0$
reads
\begin{equation}
\Delta_T U_k = \frac{\Nf}{32 \pi^2} \left[ A^p(\zeta) \cdot \kir^2 + C^p(\zeta)
\cdot \kappa^2 \right] \phi_0^2, \quad \zeta=\frac{T}{\kir},
\end{equation}
with the temperature-dependent coefficients functions
\begin{eqnarray}
\label{eq:definingAB}
A^p(\zeta) &&= -\dfrac{1}{2} \bigintsss_0^\infty
\dfrac{d\tilde{s}}{\tilde{s}^2} 
\E^{-\tilde{s}^p}  \left[\vartheta_3\left( \dfrac{\pi}{2}, 
\E^{-\frac{1}{4\zeta^2 \tilde{s}}} \right) - 1\right], \\
C^p(\zeta) &&= -\dfrac{1}{4} \bigintsss_0^\infty \dfrac{d\tilde{s}}{\tilde{s}} 
\E^{-\tilde{s}^p}  \left[\vartheta_3\left( \dfrac{\pi}{2}, 
\E^{-\frac{1}{4\zeta^2 \tilde{s}}} \right) - 1\right]
\end{eqnarray}
that depend on the regularization scheme parameter $p$ and the rescaled 
temperature $\zeta = T/ \kir$. Both functions vanish in the zero-temperature 
limit, $A^{p},C^p|_{\zeta\to 0} =0$ for any legitimate scheme parameter $p$. 
The quadratic part of the effective potential at finite temperature can be 
expressed through these coefficients functions
\begin{eqnarray}
U^T_{\kir} =&& -\dfrac{\Nf \phi_0^2}{2} \left[\dfrac{1}{\blamc} - 
\dfrac{1}{\blamL} - \dfrac{\kir^2}{16 \pi^2} \left(\Gamma \left(1-\dfrac{1}{p} 
\right)  + A^p(\zeta) \right)\right] \nonumber \\
 &&- \Nf \kappa^2 \left( 6 \xi_{\kir} - \dfrac{1}{32 \pi^2} C^p(\zeta) \right) 
\phi_0^2 + \mathcal{O}(\phi_0^4),
\end{eqnarray}
leading to the temperature-dependent curvature bound 
\begin{eqnarray}
\label{eq:Boundpz}
\dfrac{\kappa^2}{\kir^2} \leq B^p(\zeta) := \dfrac{\Gamma \left(1-\dfrac{1}{p} 
\right) + A^p(\zeta)}{192 \pi^2 \xi_{\kir} - C^p(\zeta)}.
\end{eqnarray}
This bound represents a central result of our work.
The integrals for the coefficients $A^p(\zeta)$ and $C^p(\zeta)$
can be evaluated numerically for arbitrary $p$ rather straightforwardly. For  
analytic estimates, we expand the thermal part of the heat kernel, excluding 
the zero temperature contribution, in a Taylor expansion for the second 
argument of the Jacobi theta function
\begin{eqnarray}
\label{eq:expansionJacobi}
 \left[\vartheta_3\left( \dfrac{\pi}{2}, 
\E^{-\frac{\pi^2}{(2\pi T)^2 s}} \right) - 1\right] =  2\sum_{n=1}^\infty
(-1)^n 
\ \E^{-\frac{n^2}{4\zeta^2 \tilde{s}}}. 
\end{eqnarray}
We observe that the contributions decrease exponentially for each additional 
order suggesting that expansions truncated at a certain order $N$ can still 
represent a quantitatively accurate approximation up to a 
certain temperature. Expanding the thermal coefficients from 
\Eqref{eq:definingAB} accordingly, we can express the result to all orders in 
the expansion by the two functions $a^p(z)$ and $c^p(z)$, respectively,
\begin{eqnarray}
A^p(\zeta)  &&= - \sum_{n=1}^\infty (-1)^n  \bigintsss_0^\infty 
\dfrac{d\tilde{s}}{\tilde{s}^2} \E^{-\tilde{s}^p} \E^{-\frac{n^2}{4\zeta^2 
\tilde{s}}} \nonumber \\ 
&&=: \sum_{n=1}^\infty (-1)^n \ a^p(\zeta/n),  \label{eq:Apzeta}\\
C^p(\zeta)  &&= - \dfrac{1}{2}\sum_{n=1}^\infty (-1)^n  \bigintsss_0^\infty 
\dfrac{d\tilde{s}}{\tilde{s}} \E^{-\tilde{s}^p} \E^{-\frac{n^2}{4\zeta^2 
\tilde{s}}} \nonumber \\ 
&&=: \sum_{n=1}^\infty (-1)^n \ c^p(\zeta/n) \label{eq:Cpzeta}.
\end{eqnarray}
These functions can be computed analytically for the 
scheme parameters $p=1$ and $p=\infty$ and yield
\begin{eqnarray}
a^{p=1}(z) &&= -8zK_1 \left( \frac{1}{z} \right), \\
a^{p=\infty}(z) &&= -8z^2 \E^{-\frac{1}{4z^2}} ,\\
c^{p=1} (z) &&= -2K_0 \left( \frac{1}{z} \right) ,\\
c^{p=\infty} (z) &&=\text{Ei} \left( -\frac{1}{4z^2} \right),
\end{eqnarray}
with $K_n(z)$ being the modified Bessel functions of the second kind, and 
$\text{Ei}
(z)$ the exponential integral.
Whereas the choice $p=1$, corresponding to the Callan-Symanzik regulator, is 
insufficient for regularizing the quantum fluctuations as discussed above, 
there is no problem using it for the thermal part. While setting $p=2$ for 
the quantum and $p=1$ for the thermal fluctuations does not correspond to a 
fully consistent regularization scheme, the comparison between 
$p=\infty$ and the ``$p=1,2$'' scheme can can be used for analytical 
estimates of the scheme dependence. 
A full numerical comparison between the extreme choices $p=2$ and $p=\infty
$ is shown in Fig. \ref{fig:nump1Inf}. Here, the bound $B^p(\zeta)$ of 
Eq.~\eqref{eq:Boundpz} is shown as a function of rescaled temperature for the 
two schemes. While there is a quantitative difference for low temperatures 
which reflects the scheme dependences found in Eq.~\eqref{eq:BoundT0} for 
$T=0$, this difference significantly weakens for increasing temperature. This 
enhances the predictivity of our quantitative estimates for the 
finite-temperature case.
A fully analytical estimate is obtained by truncating the series in 
Eqs.~\eqref{eq:Apzeta} and \eqref{eq:Cpzeta} at a finite order in $N$ using, 
say,  the $p\to\infty$ scheme. In Fig.~\ref{fig:numpinftyO135}, we compare 
increasing orders for $N=1,3,5$ with the corresponding full numerical result.
We observe that already low-order estimates reflect the full behavior
qualitatively rather well. For increasing order, also the quantitative precision
increases. For instance, for $N=15$ no difference between the analytical
estimate and the numerical result would be visible in Fig.~
\ref{fig:numpinftyO135} in the shown regime of rescaled temperatures as large 
as $\zeta=100$. The large-$\zeta$ behavior of the bound fits well to quadratic 
increase. A numerical fit yields $B^p(\zeta)\simeq 0.02 \zeta^2$ for the leading 
high-temperature behavior. This matches also with the qualitative behavior of 
the large-temperature expansion of the heat-kernel.
For the application of our curvature bound to a quantum gravity 
scenario below, we simply use the analytical estimate $B^{p\to\infty}(\zeta)$ 
for $N=15$, as it is sufficiently accurate for all values of rescaled 
temperature $\zeta$ of interest.

\section{Asymptotically Safe Gravity: from curvature bound to matter bound}
\label{sec:AS}

\begin{figure}[H]
\includegraphics[width= \columnwidth]{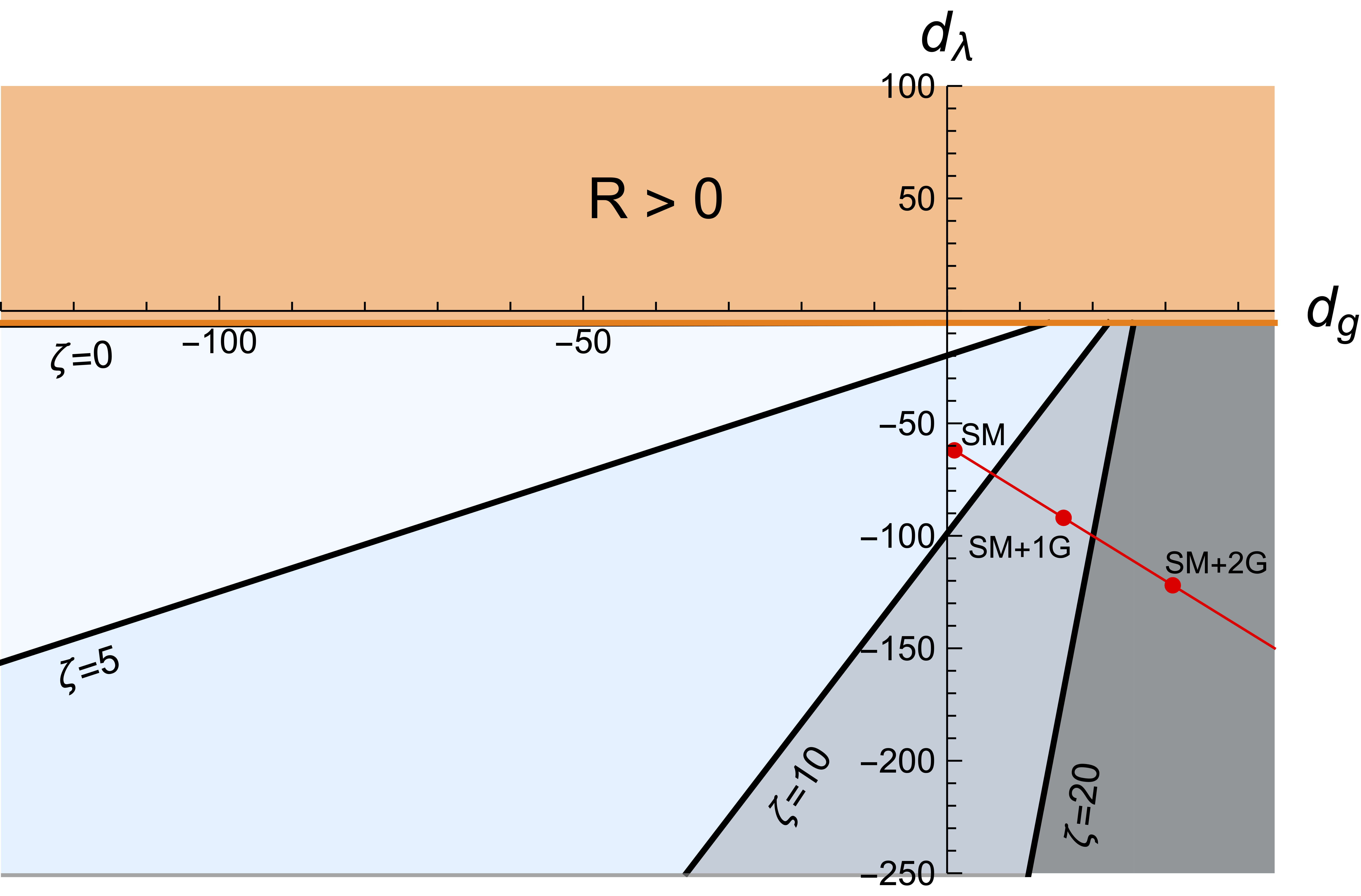}
\caption{Space of asymptotically-safe quantum gravity theories with 
matter parametrized by $d_g$ and $d_\lambda$ according to Eq.~
\eqref{eq:dgdlambda}. The orange area corresponds to regions with 
positive curvature. Each of the four solid lines distinguish regions free from 
gravitational catalysis (region above/left of each line) from regions that 
violate our curvature bound and could feature chiral 
symmetry breaking through gravitational catalysis (darker shaded region 
below/right of each line) -- for the rescaled temperatures $\zeta=0$ 
(barely visible in the upper left sector), $5$, $10$, and $20$. The red dot 
marks the Standard Model (SM) matter content with the red line indicating the 
Standard Model with additional fermionic generations.}
\label{fig:mattertheories}
\end{figure}

The preceding results can be applied to generic quantum gravity scenarios as 
soon as they feature an effective metric based description below a certain 
high-energy scale. For this, we assume that such a scenario provides 
information about the effective spacetime structure at short-distance scales, 
e.g., in the form of a possibly scale-dependent effective metric, $\langle 
g_{\mu\nu}\rangle_k$. In addition, we assume that the quantum gravity
scenario also accommodates a model of the cosmological evolution going along 
with a scale-dependent evolution of the temperature. 
In fact, the asymptotic-safety scenario for quantum gravity 
\cite{Weinberg:1976xy,Weinberg:1980gg,Reuter:1996cp,Dou:1997fg} has 
witnessed 
rapid progress over the past two decades, as, e.g., reviewed in 
\cite{Niedermaier:2006wt,Litim:2011cp,Reuter:2012id,Percacci:2017book,%
Pereira:2019dbn,Reuter:2019byg,Bonanno:2020bil,Dupuis:2020fhh,Reichert:2020mja,%
Pawlowski:2020qer}, and is thus able to provide us with required estimates 
also including matter degrees of freedom  
\cite{Percacci:2002ie,Codello:2008vh,Dona:2012am,Meibohm:2015twa,%
Christiansen:2017cxa,Eichhorn:2016vvy,Hamada:2017rvn,Eichhorn:2017eht,%
%Biemans:2017zca,Alkofer:2018baq,Eichhorn:2018yfc,Franchino-Vinas:2018gzr,%
deBrito:2019umw,Burger:2019upn,Daas:2020dyo,Eichhorn:2020sbo,deBrito:2020dta}. 
The scenario therefore serves as an example in the following. Let us briefly 
summarize the 
corresponding line of argument developed in \cite{Gies:2018jnv}, generalizing 
it to the presence of finite temperature during a cosmological evolution.  
For simplicity, we work in the so-called Einstein-Hilbert truncation, assuming 
that higher-order curvature operators -- though relevant for a more accurate 
picture of the UV 
behavior\cite{Lauscher:2002sq,Codello:2006in,Codello:2007bd,Codello:2008vh,%
Benedetti:2009rx,Falls:2013bv,Falls:2014tra,Gies:2016con,%
Denz:2016qks,Christiansen:2016sjn,Falls:2017lst,Eichhorn:2018akn,%
Falls:2018ylp,Falls:2020qhj} 
-- 
do 
not take a strong influence on the RG trajectory at the effective scales 
considered here. 
Incidentally, this approximation could straightforwardly be improved, e.g., by 
considering trajectories as in \cite{Gubitosi:2018gsl}. The effective 
scale-dependent metric obeys the quantum equation of motion which -- on the 
Einstein-Hilbert level -- corresponds to Einstein's equations,
\begin{align}
R_{\mu \nu} (\langle g \rangle_k ) - \frac{1}{2} R(\langle g \rangle_k )  
\langle g_{\mu \nu}\rangle_k +
\bar{\Lambda}_k \langle g_{\mu \nu}
\rangle_k=0
\label{eq:Einstein}
\end{align}
Within the asymptotic-safety scenario, the dimensionless version 
of the scale-dependent cosmological parameter $\bar{\Lambda}_k$ is governed 
by the Reuter fixed point, i.e., a non-Gau\ss{}ian UV fixed point 
$\lambda_\ast$, in the trans-Planckian region of the RG flow. Even though 
typical RG 
trajectories appear to spiral around the fixed point towards the UV, i.e., 
quantitatively relevant values potentially oscillate about $\lambda_\ast$ during 
the course of the RG evolution, we use this fixed-point value as an estimate
for the effective curvature of local spacetime patches averaged over a length 
scale $\sim 1/\kir$. 
Since the background $S^1\otimes\mathbb{H}^{3}$ chosen for our finite 
temperature analysis is not a solution to the Einstein equation 
\eqref{eq:Einstein}, i.e. it is not of Friedman-Lemaître-type, we cannot 
unambiguously link our background-curvature parameter $\kappa$ to the 
fixed-point value $\lambda_\ast$ of the asymptotic-safety scenario. In the 
following, we use the trace of the Einstein equation, which yields in the fixed 
point regime:
\begin{align}
\dfrac{R}{\kir^2} = 4 \lambda_\ast.
\label{eq:Einsteintr}
\end{align}
Alternatively, we could use solely the spatial components of the Einstein 
equation for which $\mathbb{H}^{3}$ is a solution; in this case, a factor of 6 
would replace the factor of 4 on the right-hand side of 
Eq.~\eqref{eq:Einsteintr}, mildly modifying our quantitative results below. In 
the following, we use the trace prescription leading to
Eq.~\eqref{eq:Einsteintr}, as it implements isotropy on the level of the 
equation of motion.  
By means of this relation, the asymptotic-safety scenario relates the curvature 
of local 
spacetime patches in the trans-Planckian regime to the fixed-point value of the
cosmological parameter. In those regimes where the latter is positive
our curvature bounds are irrelevant, as they are automatically fulfilled. 
Hence, we concentrate on the case where $\lambda_\ast<0$, for which we 
obtain an estimate for our curvature parameter:
\begin{align}
\dfrac{\kappa^2}{k^2} =  \dfrac{2|\lambda_\ast|}{3} > 0, \quad \text{for}\,\, 
\lambda_\ast<0.
\end{align}
A crucial observation within the asymptotic-safety scenario is 
that the fixed-point properties depend on the matter content 
\cite{Percacci:2002ie,Dona:2012am,Eichhorn:2018yfc}, i.e., on the 
nature of the fluctuating quantum degrees of freedom coupling to gravity. In 
the present setting, the dependence of $\lambda_\ast$ on this matter content 
comes in through two parameter combinations: 
\begin{align}
d_g = N_\text{S} - 4N_\text{V} +2\Nf, \qquad d_\lambda = N_{\text{S}} 
+2N_\text{V} -4\Nf,
\label{eq:dgdlambda}
\end{align}
where $N_\text{S}$ counts the number of scalar degrees of freedom, $N_
\text{V}$ denotes vector degrees of freedom, and $\Nf$ is the flavor number as 
before. (Here, we quote results for the so-called type IIa regulator 
\cite{Percacci:2017book} which accounts for the appropriate endomorphisms of 
the Laplacians for particles with spin \cite{Dona:2012am}). The precise 
dependence of $\lambda_\ast$ on these matter parameters is not yet fully 
determined. Current results show some dependence on the details of the 
non-perturbative approximation, see e.g. 
\cite{Percacci:2017book,Dona:2012am,Meibohm:2016mkp,Biemans:2017zca}. A 
quantitative comparison 
concerning gravitational catalysis in the zero-temperature limit can be found 
in \cite{Gies:2018jnv}. Roughly speaking,  $\lambda_\ast$ in simple 
approximations is proportional to $d_\lambda$, such that a dominant number 
of fermion flavors $\Nf$ moves the system towards the region where 
gravitational catalysis could become relevant.
For the following quantitative discussion, we use the fixed-point results of 
\cite{Biemans:2017zca} and their dependence on $d_g$ and $d_\lambda$ as 
an example. We focus on the regularization scheme $p\to\infty$, and -- unless 
stated otherwise -- assume the scalar-curvature coupling at its conformally 
coupled point $\xi_{\kir}=1/6$ which is known to be a fixed-point of the 
universal part of the perturbative RG 
\cite{Buchbinder:1992rb,Shapiro:2015ova,Merzlikin:2017zan}. In order to 
complete the concrete scenario of our investigation, 
we need to connect the scale $\kir$ at which we consider the system with a 
value (or range of values) for the temperature $T$. In a specific cosmological 
model, the temperature would be connected with a relevant cosmological scale, 
say, a time parameter or an expansion scale. Within asymptotically-safe 
cosmologies, such scales are assumed to be linked to some suitable power of $k
$ by RG-improvement arguments 
\cite{Bonanno:2001hi,Bonanno:2001xi,Guberina:2002wt,Reuter:2005kb,%
%Bonanno:2005mt,Weinberg:2009wa,Cai:2011kd,Copeland:2013vva,Saltas:2015vsc,%
Bonanno:2015fga,Bonanno:2017pkg,Bonanno:2018gck,Platania:2019qvo,%
Platania:2020lqb}. In fact, several scale-setting procedures have been discussed 
in the literature  
\cite{Bonanno:2001xi,Babic:2004ev,Platania:2020lqb,Koch:2014joa}. For 
the present study, we therefore use the 
rescaled temperature $\zeta=T/\kir$ as a parameter, the value (or range of 
relevant values) will be fixed by a specific choice of the cosmological model. 
Simple RG-improvement arguments suggest to consider $\zeta\sim \mathcal{O}
(1)$.
Since or zero-temperature bound on $\mathbb{R}\otimes H^3$ is quantitatively 
stronger than the corresponding one on $H^4$ for $\xi_{\kir}=0$ as used in 
\cite{Gies:2018jnv}, we expect a correspondingly larger extent of the regime 
where gravitational catalysis could be active. Given our result that the 
curvature bound \eqref{eq:Boundpz} weakens for increasing temperature, the 
region which is not affected by gravitational catalysis should increase with 
$\zeta$. In fact, this is visible in Fig.~\ref{fig:mattertheories}: here the 
orange region in the upper part of the plot indicates the region where 
$\lambda_\ast$ is positive in the asymptotic-safety scenario, hence this region 
is not affected by gravitational catalysis. 

At finite rescaled temperature $\zeta=T/\kir$, the solid lines separate the 
regions in this space of asymptotically safe theories with matter which are free 
of gravitational catalysis (regions above/left of lines) from those where  
our curvature bound is violated and gravitational catalysis could trigger 
fermion mass generation (darker shaded regions below/right of lines). In fact, 
the curvature bound for $\zeta=0$ is rather close to the $R>0$ curve with only 
a slim unaffected region extending along the negative $d_g$ axis (hardly visible 
on the scale of this Fig.~\ref{fig:mattertheories}). This agrees with the 
comparatively strong curvature bound on $\mathbb{R}\otimes H^3$ for 
$\xi_{\kir}=1/6$ and should be taken as an indication that gravitational 
catalysis might be more relevant than previously anticipated for the $H^4$ 
background. In other words, the details of the spacetime structure of local 
spacetime patches do matter beyond the simple statement of positive or negative 
average curvature and thus need to be addressed by the quantum-gravity 
scenario under scrutiny.

For increasing rescaled temperature $\zeta$ the region 
satisfying the curvature bound increases; for $\zeta> \mathcal{O}(10)$, the 
boundary line approaches a vertical line that ultimately matches with a region 
where the computation of \cite{Biemans:2017zca} does no longer find a viable 
UV fixed point. 

It is interesting to observe that the standard model (SM) matter content with 
three generations and thus $N_\text{S}=4$, $N_\text{V}=12$ and $\Nf=45/2$ 
(excluding right-handed neutrino components) (red dot in 
Fig.~\ref{fig:mattertheories}) lies in the region violating the bound for small 
$\zeta$ but satisfying the bound for $\zeta>8.3$ for the 
current assumptions. This illustrates directly that a given quantum gravity 
scenario does not automatically allow for an arbitrary matter content. 
Depending on the details of the local spacetime curvature, gravitational 
catalysis could be relevant and needs to be carefully scrutinized in this 
regime.

At the same time, our current study also reveals, how gravitational catalysis 
endangering the existence of light fermions could be tamed in the course of the 
cosmological evolution: even a critical spacetime curvature violating the 
zero-temperature bound may not give rise to gravitational catalysis and fermion 
mass generation provided the temperature remains sufficiently high compared to 
the averaging scale $\kir$. From an RG perspective, this can be understood in 
terms of the thermal masses of the fermions, which effectively suppress the 
fermionic fluctuations. This inhibits the symmetry breaking channels to become 
RG relevant as predicted by zero-temperature catalysis. A similar mechanism has 
been investigated in scenarios of Higgs inflation in order not to be affected 
by further minima in the Higgs potential \cite{Bezrukov:2014ipa}.

This argument can also be inverted: in order to evading gravitational catalysis 
for a given matter content in asymptotically safe gravity, the cosmological 
evolution in the early universe has to go along with a sufficiently high 
(rescaled) temperature. In this way, gravitational catalysis can put bounds on 
the cosmological model. 

As we parametrize such models using the rescaled temperature, a given 
value of $\zeta$ -- which should be understood as a lowest value in a given 
model in the early universe -- can accommodate a certain matter content. In 
order to illustrate this dependence, we concentrate on standard-model-like 
theories possibly with extra generations of fermions. In 
Fig.~\ref{fig:mattertheories}, these theories move along the red line towards 
the region increasingly endangered by gravitational catalysis with the cases of 
additional complete generations (``+1G'', ``+2G'') marked by red dots. 

\begin{figure}[H]
\includegraphics[width= \columnwidth]{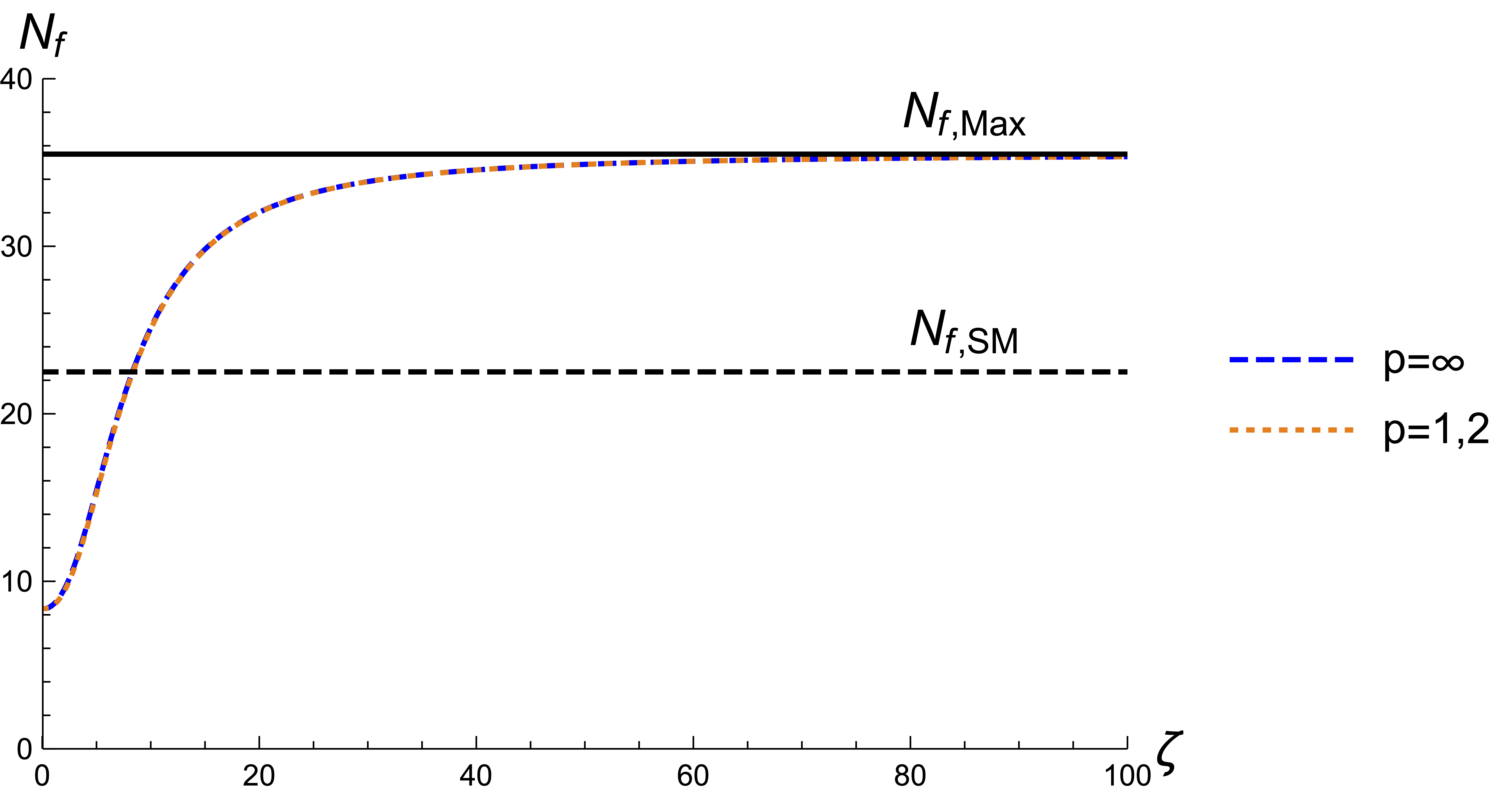}
\caption{Number of fermion species for a standard-model-like particle 
content ($N_{\text{S}}=4, N_{\text{V}}=12$) compatible with the curvature bound 
from gravitational catalysis as a function of the rescaled temperature $\zeta$ 
for different regularization schemes $p$ and the scalar-curvature coupling 
$\xi_{\kir} = 1/6$. The solid black line represents the upper bound 
$N_{\text{f, Max}}=35.5$ which is approached in the limit $\zeta \rightarrow 
\infty$; the dashed line marks the number of fermions in the standard model 
$N_{\text{f, SM}}=22.5$.}
\label{fig:Nfzetap}
\end{figure}

By virtue of the fixed-point structure, an arbitrarily large 
number of fermions is not supported. This is visible in Fig.~\ref{fig:Nfzetap}, 
where the allowed number of femions $\Nf$ compatible with our bound is plotted 
as a function of $\zeta$. The observed threshold set by the standard-model 
fermion content is marked by a horizontal dashed line; it is surpassed for 
$\zeta>8.3$. Even at asymptotic temperatures, a maximum fermion number of 
$\Nf{}_{\text{max}} = 35.5$ is approached. In 
this 
figure, we also illustrate the scheme dependence of our finite 
temperature results by showing the extremal parameter choices $p\to\infty$ and 
the mixed approximate scheme $p=1,2$. On the scale of this figure, hardly any 
variation is recognizable, which illustrates that the scheme-dependencies are 
under control here. 

\begin{figure}[H]
\includegraphics[width= \columnwidth]{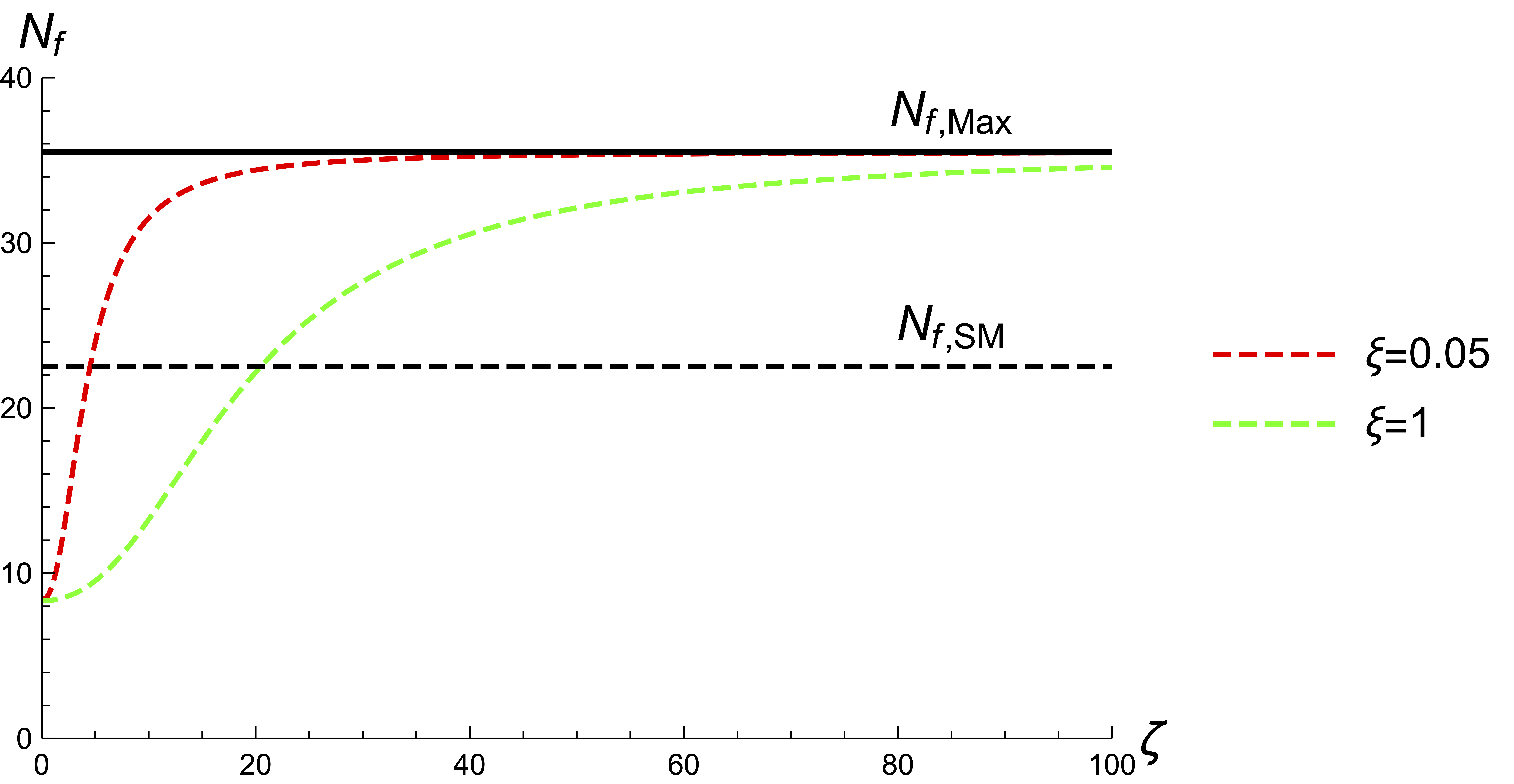}
\caption{Number of fermion species for a standard-model-like particle 
content ($N_{\text{S}}=4, N_{\text{V}}=12$) compatible with the curvature bound
from gravitational catalysis as a function of the rescaled temperature $\zeta$ 
for different scalar-curvature coupling parameters $\xi_{\kir}$ using the 
regularization scheme $p=\infty$. The horizontal lines are as in
Fig.~\ref{fig:Nfzetap}.}
\label{fig:Nfzetaxi}
\end{figure}

By contrast, there is a stronger dependence on the  
scalar-curvature coupling $\xi_{\kir}$. Nevertheless, while the 
zero-temperature bound is inversely proportional to and 
thus rather strongly varying with $\xi_{\kir}$, the finite-temperature results 
are somewhat less sensitive. This is visible in Fig.~\ref{fig:Nfzetaxi}, where 
the number of fermions $\Nf$ that can be accommodated is shown for 
$\xi_{\kir}=0.05$ and $\xi_{\kir}=1$. Both curves eventually surpass the 
standard-model threshold, however for different values of the rescaled 
temperature.  

In summary, the asymptotic-safety scenario for quantum gravity together with 
standard-model matter content can evade the curvature bound imposed by 
gravitational catalysis provided the temperature is sufficiently high in the 
course of the cosmological evolution. By contrast, theories with a more 
dominant fermionic matter content either require much higher temperatures to 
comply with the bounds or fail to support a UV-completing fixed point.

\section{Conclusions}
\label{sec:conc}

The present work generalizes the concept of curvature bounds from gravitational 
catalysis \cite{Gies:2018jnv} to finite temperatures as well as to the case of 
a spatially curved space. In addition to the role played by the thermal 
effects, we observe that the details of the averaged curvature of local patches 
of spacetime matters significantly: First, gravitational catalysis is more 
strongly triggered for the spacetime $\mathbb{R}\otimes H^3$ than for the 
maximally symmetric case $H^4$. Second, also the dependence on the 
scalar-curvature coupling $\sim \xi \phi^2 R$ is much more prominent for the 
former case than for the latter. Both observations have a strong influence on 
the curvature bound that indicates how the details of the curvature background 
matter for the phenomenon of gravitational catalysis. 

It is important to emphasize that the curvature bound derived in this paper is 
an estimate for the extent of the region that is not affected by gravitational 
catalysis according to our assumptions. If a system (e.g., subject to a 
specific quantum gravity scenario) violates the curvature bound, this does not 
necessarily imply that fermion mass generation kicks in as a manifestation of 
gravitational catalysis. Further dynamical mechanisms could still avoid the 
occurrence of gravitational catalysis. For instance, fluctuations of the 
scalar order parameter tend to weaken the symmetry-breaking channel. On the 
other hand, a finite initial fermionic self-interaction could enhance the 
tendency towards fermionic gap formation. Also, the curvature bound does not 
account for the possibility of further local minima which could become the 
global one at a first-order transition; our method is only sensitive to 
second-order transitions. Of course, first-order transitions could 
straightforwardly be detected by a global study of the effective potential. If 
they occur, they would strengthen our bounds. 

As a first example, we have applied the curvature bound to the 
asymptotic-safety scenario for quantum gravity. A rather robust prediction of 
this scenario that relies on the existence of an interacting UV-fixed point is 
that the cosmological constant can have a negative sign in the short-distance 
regime (with a dynamical transition to positive values for the long-range 
physics) depending on the matter content. In particular, a dominance of 
fermionic matter degrees of freedom pushes the fixed point of the 
cosmological term to negative values. RG-improvement arguments then suggest 
that the properties of the quantum spacetime in the short-distance regime can 
effectively be described by a scale-dependent version of Einstein's equations 
(or higher-derivative versions thereof). For our purposes this suggests that 
local patches of spacetime appear as effectively negatively curved. If so, this 
effective negative curvature also enhances the symmetry-breaking channels of 
fermionic fluctuations by (the scale-dependent version of) gravitational 
catalysis. If symmetry breaking was triggered in the high-energy regime of 
gravity, fermions would acquire a mass proportional to the scale of symmetry 
breaking. Gravitational catalysis would therefore inhibit the existence of 
light fermions in Nature.

This line of argument thus connects the observational fact of light fermions 
with properties of quantum spacetime in the high-energy regime. By extending 
the RG-improvement argument to a cosmological setting, our reasoning connects 
the curvature bound of gravitational catalysis to a combination of matter 
degrees of freedom such as the fermion flavor number together with the thermal 
evolution of the universe, parameterized in this work by the 
rescaled temperature $\zeta$. 

Whereas our results for the general curvature bound have a clear quantitative 
meaning within the given assumptions, the application to the asymptotic safety 
scenario should be considered as more qualitative because of the approximations 
involved and the genuine qualitative nature of RG improvement. Therefore, we 
interpret these results as an indication that the asymptotic-safety scenario 
for quantum gravity can indeed be compatible with the existence of light 
fermions; there is definitely room for evading the bounds imposed by 
gravitational catalysis for particle models with a matter content similar to 
that of the standard model. While our line of argument based on 
gravitational catalysis can put an upper bound on the number of fermionic 
degrees of freedom, it is interesting to see that a combination of gravity and 
abelian gauge interactions of fermions can also produce a lower bound 
\cite{deBrito:2020dta}.

We believe that it will be highly worthwhile to 
check for the role of gravity, specifically gravitational catalysis, and the 
consistency with light 
fermions in other scenarios of quantum gravity as well. While our bounds can be 
applied in other settings, the inclusion of matter degrees of freedom is a 
common effort in many research directions of quantum gravity 
\cite{Oriti_2002,Bianchi_2013,Steinhaus:2015kxa,Catterall:2018dns,% 
Glaser:2018jss,Ambjorn:2021fkp}.

%%%%%%%%%%%%%%%%%%%%%%%%%%%%%%%%%%%%%%%%
\section*{Acknowledgments}
%%%%%%%%%%%%%%%%%%%%%%%%%%%%%%%%%%%%%%%%

We thank Astrid Eichhorn, Aaron Held, Riccardo Martini, Martin Pauly, Alessia 
Platania and Marc Schiffer
for valuable discussions. We are grateful to Frank Saueressig for making his data 
and Mathematica files available to us. This work has been funded by the 
Deutsche Forschungsgemeinschaft (DFG) under Grant No. 406116891 within the 
Research Training Group RTG 2522/1.
% \\
% \appendix
% 
% \section{Heat kernels or some other stuff}
% \label{app:heatKer}
% 
% Test
% 
% \section{Foliated Spacetime Flow Equations}
% \label{app:AS}
% 
% Test

\bibliography{bibliography}

\end{document}